# Team formation and team performance: The balance between team freshness and repeat collaboration

Meijun Liu[1], Ajay Jaiswal[2], Yi Bu[3], Chao Min[4], Sijie Yang[1], Zhibo Liu[3], Daniel Acuña[5], Ying Ding[2*]

1 Institute for Global Public Policy, Fudan University, Shanghai, China.
2 School of Information, The University of Texas at Austin, Austin, TX, U.S.A.
3 Department of Information Management, Peking University, Beijing, China.
4 School of Information Management, Nanjing University, Nanjing, China.
5 School of Information Studies, Syracuse University, Syracuse, NY, U.S.A.

*Corresponding author: E-mail: ying.ding@austin.utexas.edu

Keywords: team formation; team performance; freshness; repeat collaboration

**Abstract**

Incorporating fresh members in teams is considered a pathway to team creativity. However, whether freshness improves team performance or not remains unclear, as well as the optimal involvement of fresh members for team performance. This study uses a group of authors on the byline of a publication as a proxy for a scientific team. We extend an indicator, i.e., team freshness, to measure the extent to which a scientific team incorporates new members, by calculating the fraction of new collaboration relations established within the team. Based on more than 43 million scientific publications covering more than a half-century of research from Microsoft Academic Graph, this study provides a holistic picture of the current development of team freshness by outlining the temporal evolution of freshness, and its disciplinary distribution. Subsequently, using a multivariable regression approach, we examine the association between team freshness and papers' short-term and long-term citations. The major findings are as follows: (1) team freshness in scientific teams has been increasing in the past half-century; (2) there exists an inverted-U-shaped association between team freshness and papers' citations in all the disciplines and in different periods; (3) the inverted-U-shaped relationship between team freshness and papers' citations is only found in small teams, while, in large teams, team freshness is significantly positively related to papers' citations.

# 1 Introduction

Teamwork is increasingly pervasive and drives innovations in the contemporary scientific landscape (Liu, Wu, Rousseau, & Rousseau, 2020; Milojević, 2014). Compared to singular knowledge production, collaborative work can yield more discoveries and breakthroughs (Liu et al., 2020; Wuchty, Jones, & Uzzi, 2007). Despite considerable efforts to explore the heterogeneity of team members, we know little about how teams are formed to incorporate fresh members who have no prior collaboration with other team members, and incumbents who have repeat collaborative connections with other team members, and the relationship between team freshness and team performance.

The evolution of team freshness might be accompanied by historical changes in science. Contemporary science has witnessed an exponential growth of scientific publications and a rise of the volume of the scientific labor force (Baskaran, 2017; Fortunato et al., 2018; Price, 1963). Scientific research is characterized by a fundamental shift toward team-based research (Rawlings & McFarland, 2011; Wuchty et al., 2007). Due to the increasing specialization in science (Cole & Harriet, 2017; Evans, 2016; Jones, 2009; Moody, 2004), the need to combine diverse and interdisciplinary knowledge and skills to address complex research problems (Katz & Martin, 1997) and the growing costs of scientific facilities and instruments (Shrum, Genuth, Carlson, & Chompalov, 2007), in almost all branches of science, scientists are increasingly involved in teamwork (Wuchty et al., 2007). In addition, advances in information and communication technologies (Binz-Scharf, Kalish, & Paik, 2015) and reductions in travel costs (Katz & Martin, 1997), have made constructing new collaboration links easier. These changes have led to the increasing need to involve new members in teams for diverse and distant knowledge, reduced the costs of constructing fresher teams, and might change the development of team freshness. To understand the temporal changes in team freshness from a historical standpoint, we propose **RQ1. How has team freshness evolved in the past decades?**

Fresh members enhance team learning and thus spur team creativity by bringing new knowledge and novel perspectives (Perretti & Negro, 2007; Rosenkopf & Almeida, 2003; Skilton & Dooley, 2010). Teams might become less creative over time due to groupthink, homogeneity, and less tendency to disturb the status quo (West & Anderson, 1996). Entry of fresh members could generate conflicts and divergent opinions that can trigger creativity (Badke-Schaub, Goldschmidt, & Meijer, 2010; Farh, Lee, & Farh, 2010; Kane, Argote, & Levine, 2005; Santos, Uitdewilligen, & Passos, 2015). In this sense, team freshness could lead to high team performance. A recent study found that team freshness is positively related to a paper's originality and multi-disciplinary impact, which is consistent with this argument (Zeng, Fan, Di, Wang, & Havlin, 2021).

However, too much team freshness could be risky and harmful to team performance due to the high cost of forming new ties (Jackson, Stone, & Alvarez, 1992) and fresh members' adaption (Chen, 2005), and less trust and familiarity (Van Der Vegt, Bunderson, & Kuipers, 2010). From the psychological perspective, repeat collaboration entails greater certainty, trust, and reciprocity, and more efficient knowledge transfer, all of which offsets the negative consequences of team freshness and thus facilitates team performance (Dahlander & McFarland, 2013; Uzzi, 1997). However, high levels of repeat collaboration could dampen the creation of innovative ideas by reducing collaboration efficiency, homogenizing the pool of knowledge, narrowing search spaces of teams, and reducing conflicts (Guimera, Uzzi, Spiro, & Amaral, 2005; Porac et al., 2004). Due to the co-existing benefits and disadvantages of team freshness and repeat collaboration, empirical evidence shows that the combination of team freshness and repeat collaboration leads to the best team performance (Guimera et al., 2005; Perretti & Negro, 2007). The mixed arguments about the positive and negative consequences of freshness and repeat collaboration prompt us to explore the "bliss point" between team freshness and repeat collaboration that brings the optimal team performance. Thus, we raise **RQ2. What is the relationship between team freshness and team performance?**

The relationship between team freshness and team performance might vary with changes in team size. Whether and how team freshness shapes team performance relies on whether benefits and detriments caused by team freshness outweigh each other, which might be different across team size. Fewer and simpler links are embedded in small teams where negative impacts caused by a high level of team freshness could disrupt the whole team (Zhang, Huang, Shao, & Huang, 2020) and thus worsen team performance, while negative impacts of team freshness might be subtle in large teams and so barely influence team performance. A few empirical studies suggest that stability is more important for small teams' survival, and that large teams gain benefits from membership dynamics (Palla, Barabási, & Vicsek, 2007; Zhang et al., 2020). However, it is still unclear whether and how the relationship between team freshness and team performance is shaped by team size. Thus, we propose **RQ3: Is the relationship between team freshness and team performance in small teams different from that in large teams?**

To address the three research questions, based on more than 43 million articles between 1951 and 2018 from Microsoft Academic Graph, the research objectives of this paper are threefold: to provide a comprehensive introduction to the evolution of team freshness, to explore how freshness is related to team performance, and to investigate whether and how the relationship between team freshness and team performance depends on team size. This study contributes to the existing literature in multiple dimensions. A better understanding of the balance between freshness and repeat collaboration in teams could improve our knowledge of how the combination of team members' characteristics relates to team outcomes from a perspective of dynamic team formation. In addition, turnover of members that involves the arrival of new members or the departure of incumbents is increasingly common in scientific teams, which can have profound consequences for team performance by altering the distribution of knowledge and skills within teams, and the relations among team members (Levine, Moreland, Argote, & Carley, 2005). From a practical perspective, the investigation into the balance of freshness and repeat collaboration sheds light on how teams achieve the best performance by maintaining a certain proportion of incumbents and absorbing some fresh members. The remainder of the paper is organized as follows. The related work section reviews the current state of the art, and is followed by section 3, where data and methodology are introduced. Section 4 presents the results. The last section discusses findings and implications for relevant policies.

# 2 Related Work

In this section, we review three strands of literature concerning the three research questions. We review the literature on team freshness and repeat collaboration, and further provide a discussion on how team size might shape the relationship between team freshness and team performance.

## 2.1 Team freshness

Team freshness could increase a team's knowledge stock (Kane et al., 2005), enhance team learning, and thus facilitate the generation of new ideas (Farh et al., 2010). Expanding the scope and space of searching and reconfiguration of diverse resources from distant sources is important for team performance and innovation (Rosenkopf & Almeida, 2003; Rosenkopf & Nerkar, 2001; Smith & Todd, 2005). Fresh members who bring new and diverse ideas, skills, and opinions could broaden the "search space" of teams, and stimulate teams to consider new ideas, adopt new practices, and improve innovation (van de Water, Ahaus, & Rozier, 2008). A few theoretical and empirical studies have provided evidential support for the promoting effect of freshness on team performance (Arrow & McGrath, 1993; Ziller, Behringer, & Goodchilds, 1962). Using the data on 6,446 movies produced by the Hollywood studio, Perretti and Negro (2007) found that the introduction of newcomers in teams leads to the genre innovation of films. Guimera et al. (2005) constructed a model for the self-assembly of creative teams and considered the fraction of newcomers in teams an important parameter. Analyzing more than 0.48 million scientific papers in physics, a recent study suggests that teams with a higher level of freshness, i.e., more fresh members are involved, are more likely to produce more original research papers (Zeng et al., 2020). Additionally, freshness could improve team performance by generating group conflicts that could facilitate team creativity. Moderate levels of task conflicts in teams could nurture creativity as, when team members disagree about central parts of the task, they will search for opposing solutions and discuss different opinions, which facilitates the creation of innovative ideas (Badke-Schaub et al., 2010; Farh et al., 2010).

Despite the benefits of freshness, it could also be a dark side for team performance due to social and psychological reasons. Stability and continuity of personal relationships are essential for the development of social integration within teams (Kane & Rink, 2015), which is important for team performance. Familiarity among team members could smooth communication and coordination among members, and encourage positive and team-building behavior (Rockett & Okhuysen, 2002). However, newcomer entry might cause unfamiliarity and reduced trust, which undermines social integration in teams and has negative effects on team performance (Van Der Vegt et al., 2010). In addition, when fresh members enter a team, they must fit in with the team culture by learning about the team's goals and values, routines and rituals, which entails time costs (Cooper, Rockmann, Moteabbed, & Thatcher, 2021). Additionally, newcomer entry will also weaken feelings of a shared team identify and introduce uncertainty (Van Der Vegt et al., 2010), which impedes the improvement of team performance. Experiencing a liability of foreignness, fresh members have disadvantages in higher information search costs, stereotyping, and marginalization by incumbents (Jackson et al., 1992). The above negative consequences caused by fresh member entry are harmful to team performance.

In short, current literature suggests two major mechanisms by which team freshness could improve team performance, i.e., bringing a broader scope of resources teams can access, and generating conflicts in teams that facilitate team creativity. At the same time, high levels of team freshness are considered harmful to team performance because of social and psychological reasons. However, none of the previous studies addressed how much freshness brings the best team performance. Additionally, most prior literature is based on psychology and organization science, making the relationship between team freshness and scientific team performance unclear.

## 2.2 Repeat collaboration

Researchers engage in repeat collaboration by publishing with the same set of authors repeatedly. Compared to team freshness, repeat collaboration entails fewer information transfer costs, greater mutual trust, commitment and reciprocity, higher efficiency at managing the details of executions, and more risk sharing (Axelrod & Hamilton, 1981; Cole & Bruno Teboul, 2004; Granovetter, 1985; Petersen, 2015; Skilton & Dooley, 2010). Several empirical studies have indicated the promoting effects of repeat collaboration on team performance. For example, Dahlander and McFarland (2013) found that repeat collaboration occurs when familiar people reflect on the quality of their relationship and shared experience. They demonstrated a lot of benefits brought by repeat collaboration, such as having fewer costs than forming new collaboration ties, entailing more certainty and trust, and more efficient knowledge transfer and reciprocity. These benefits could bring greater returns on the rate of productivity and performance quality. Petersen (2015) found that sustainable collaboration had a significantly positive influence on productivity and citations. Several studies have provided explanations regarding why repeat collaboration improves team performance from the perspective of psychology. Repeat collaboration facilitates the construction of team emotional intelligence, which is defined as the ability of team members to generate a set of shared norms that manage emotional processes (Druskat & Wolff, 2001). These norms could improve the cohesiveness of team members' collaboration, which is important for team efficiency (Druskat & Wolff, 2001). Evidence shows that team emotional intelligence could lead to a stronger relationship between team members (Jordan & Troth, 2004), efficient information exchange (Pelled, Eisenhardt, & Xin, 1999), and better decision making and reduced team conflict (Jehn & Mannix, 2001). These benefits facilitate the improvement of team performance.

However, repeat collaboration could worsen team performance, especially team creativity, because of myopic learning and conflict reduction. This claim has been supported by a few empirical and theoretical studies. Guimera et al. (2005) found that teams with high levels of repeat collaboration are less likely to publish in high-impact journals as repeat collaboration and shared experiences among team members narrow the scope of knowledge and thus hamper the production of innovative ideas. From a theoretical perspective, Skilton and Dooley (2010) demonstrated that mental models developed in prior collaboration experience might blind members to information or knowledge that could create new and innovative ideas because earlier collaboration might suppress disclosure of new ideas that would disturb the status quo. As a result, teams in which members have shared collaboration experience are more likely to converge on familiar objectives. Based on data from more than two million research articles in computer science, Bu et al. (2018) found that persistent scientific collaboration does not always generate high-quality papers since highly persistent collaboration cannot expand collaborators' networks and provide access to new knowledge from other peers. In addition, high levels of repeat collaboration might be harmful to team performance as incumbents in a team are the major source of inertial behavior and resistance to new solutions (Rollag, 2004). Another important mechanism by which repeat collaboration impedes team creativity is in reducing conflicts among team members that are sometimes beneficial for the generation of innovative ideas (Badke-Schaub et al., 2010; Farh et al., 2010).

In sum, previous studies provide contradictive arguments about the relationship between repeat collaboration and team performance due to the co-existing advantages and disadvantages it could bring. Team freshness and its opposite side, repeat collaboration, have their benefits and detriments, which might suggest that their relationship with team performance might not be simply linearly correlated. Based on previous studies, moderate levels of freshness and repeat collaboration in a team should make the team perform at its best, but there is no rigorous empirical investigation on the "bliss point" concerning the balance between freshness and repeat collaboration in scientific teams.

## 2.3 Team size

The advantages and disadvantages of large team size co-exist. Some believe that larger team size benefits team performance due to more resource advantages. Earlier studies indicated that larger teams can access a larger pool of resources and have a higher capacity to manage external uncertainties (Aldrich & Pfeffer, 1976; Pfeffer & Salancik, 2003). Team size is enlarged by incorporating more scientists into the team who directly bring more material resources, such as funding, staff, and equipment, and more intangible resources, such as knowledge, data, and social capital. The enlargement of team size enables scientists in a team to complement their skills and access a broader scope of resources to meet task demands (Zhu, Liu, & Yang, 2021). The information-processing model suggests that teams benefit from the resource advantage caused by larger team size because the novel solutions of complex problems require diverse and complementary resources (Harrison & Klein, 2007; Lee, Walsh, & Wang, 2015; Liu et al., 2020; Nederveen Pieterse, Van Knippenberg, & Van Dierendonck, 2013). This claim has been supported by empirical evidence (Hülsheger, Anderson, & Salgado, 2009).

In contrast, some studies demonstrate detriments caused by large team size, making it debatable whether the increase in team size brings better team performance. Larger teams entail more start-up costs of connecting and establishing relationships with others, and higher maintenance costs of communication and coordination, which might impede the effectiveness of knowledge production in teams (Badar, Hite, & Ashraf, 2015; Vasileiadou & Vliegenthart, 2009). Additionally, the negative effects of large team size are discussed from the perspective of social psychology. The potential gains from a large team size could be offset by "process losses", which might lead to worse team performance (Cummings, Kiesler, Bosagh Zadeh, & Balakrishnan, 2013). For example, it might be more challenging for members in a larger team to reach a common definition of the team's goal, effectively manage the workflow, and maintain efficient cooperation and communication (Chompalov, Genuth, & Shrum, 2002; Okhuysen & Bechky, 2009). In addition, problems concerning more free-riding (Backes-Gellner, Werner, & Mohnen, 2015), herding and group think (Lorenz, Rauhut, Schweitzer, & Helbing, 2011; Tetlock, Peterson, McGuire, Chang, & Feld, 1992), more emotional conflicts (Amason & Sapienza, 1997), and lower consensus (Manners, 1975), might be more likely to occur in larger teams. For example, Lee et al. (2015) found a curvilinear association between team size and scientific novelty and further suggested the underlying mechanism that team size influences novelty through increasing knowledge variety.

The optimal level of team freshness might depend on team size. The above discussion presents the changeable group environments caused by changes in team size. The impact of changes in membership, e.g., incorporating new members, is greater in a small team than in a large team because the network within a smaller team is simpler and any detriments could destroy the entire team (Zhang et al., 2020). This implies that the disadvantages brought by a high level of team freshness, such as management and communication problems, might impede team performance in small teams. In contrast, because of the large number of links in a large team, the negative impacts of team freshness can barely hinder such a team's performance. Analyzing co-authorship networks on physics papers, Palla et al. (2007) found that a high rate of turnover, i.e., the arrival of new members and departure of old members, could make a large team survive longer, while stability is required for the persistence of a small team.

Despite various efforts to examine how team size shapes scientific team performance, it is unclear whether the relationship between team freshness and team performance holds with different team sizes. Driven by this research gap, this study explores whether the relationship between team freshness and team performance in small teams is different from that in large teams.

# 3 Data and Methodology

## 3.1 Data

In this study, we use Microsoft Academic Graph (MAG), a heterogeneous graph comprised of 173.67 million scientific papers published between the years 1800 and 2018. Due to its comprehensive coverage, MAG has become an important resource for scholarly communication studies in recent years (Herrmannova & Knoth, 2016). We acquire the following information for papers from MAG: paper id, discipline, publication year and citation network, and author name(s), and affiliations. We associate every author in our dataset with an affiliation and identify the country in which that affiliation is located using Wiki data and Google search results parsing.[1]. We sum up the incoming citation links of each paper and obtained the papers' citation counts. In MAG, name disambiguation has been achieved using machine-learning and crowdsourcing approaches. A unique identifier for each author can be provided to the public, which allows us to capture authors' publication history and identify whether two authors have prior collaboration in a year. Papers with a single author are removed, and thus the final dataset for our analysis covers more than 43 million papers published from 1950 to 2018. The statistics of the final dataset used in our analyses are shown in Table S1 in the Supplementary Information. The distribution of disciplines in the final dataset is shown in Figure S1.

## 3.2 Methods

**Measuring a scientific team**

Following standard practice, we view the presence of more than one author on the byline of a publication as a scientific team (Larivière, Gingras, Sugimoto, & Tsou, 2015; Milojević, 2014). However, co-authorship data cannot fully represent scientific teamwork as it fails to capture informal collaboration, such as sharing of ideas and data (Lewis, Ross, & Holden, 2012), and collaborative efforts of scientists who are not listed on a scientific paper. Despite this drawback, co-authorship data is considered a visible, easily quantifiable, and reliable indication of collaborative activities (Franceschet & Costantini, 2010; Milojević, 2014), and allows us to capture the key elements of collaboration (Hara, Solomon, Kim, & Sonnenwald, 2003).

**Measuring team freshness and team performance**

Drawing on authorship data, we extend the measure of team freshness. Being listed as an author might suggest that this author made a certain level of contribution to the article, in a number of ways, e.g., providing ideas, skills, data, and funding resources (Leahey, 2016). Therefore, we assume that authors' participation in teamwork could be reflected by the appearance of their names on the byline of an article produced by the team. Following Zeng's definition (2021), we define team freshness as the fraction of new collaboration relations established within the team. The major difference between Zeng's indicator and team freshness in this study is that we consider the recency of collaboration when we identify whether or not collaboration between two authors is new collaboration in the focal paper. Interpersonal relationship tends to decay with time (Burt, 2000, 2002; Liu & Hu, 2021). If the most recent collaboration between two authors was established a long time ago, there are barely any benefits they could gain from their past collaboration and it only exerts a small influence on their current collaboration. We introduce a parameter that reflects a time window, $\delta$ (for fair comparison, we keep $\delta = 20$), to study the existence of prior collaborations between authors of a paper. For authors of a paper published in year t, if two authors have no collaboration in the previous 20 years, they are viewed as an author pair with no prior collaboration.

To better illustrate the measurement, we build a collaboration network of team members prior to new collaboration in the focal paper. We count the number of possible missing links in the prior collaboration network, as, after a new collaboration, the collaboration network will be a complete graph where every team member is connected with each other. Figure 1 presents an example of how to measure team freshness. There is a paper, $P$, published in 1986, which has been co-authored by five authors, $A_1$, $A_2$, $A_3$, $A_4$, and $A_5$. Figure 1(a) and 1(b) illustrate the collaboration network of each author of $P$ before and after collaborating on P, respectively. We identified whether an author pair has collaboration links between 1965 and 1985, and, if it has, the two authors in this pair will have a link connecting them in the prior collaboration network (Figure 1(a)). We find that the total number of missing links in the prior collaboration network, which are newly introduced in *P*, is four. The possible author pair in paper P is $C_5^2$; that is, ten. As shown in Equation 1, freshness for *P* should be 4/10=0.4. The range of freshness is between 0 and 1.

---

[1] For authors who have multiple affiliations, we associate them with their first affiliation.

$$Freshness_p = \frac{\# \text{ of Missing Links}}{\# \text{ of Total Possible Links}} = \frac{4}{10} = 0.4 \quad (1)$$

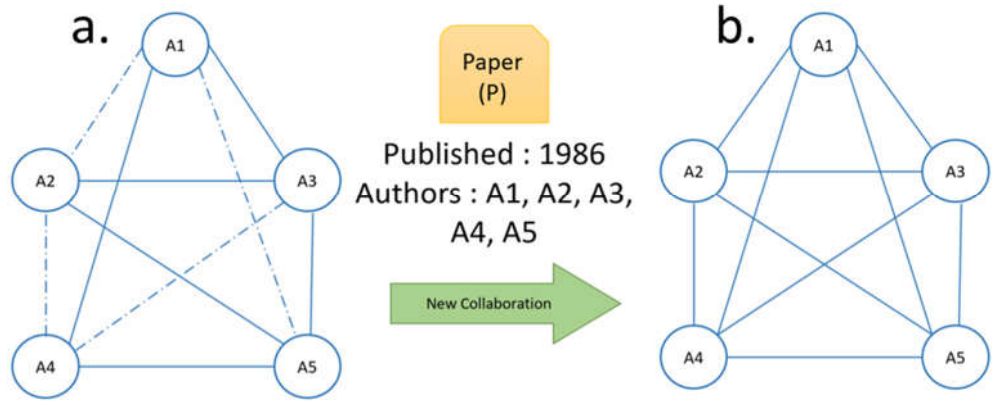

**Figure 1. An example to illustrate team freshness.** The solid lines in the network mean that two authors that are connected with each other have prior collaboration relationship. The dashed lines indicate that two authors that are connected with each other have no prior collaboration relationship.

To address RQ2 and RQ3, we use citation counts in a two- and ten-year time window to measure team performance. Team performance in science could be measured from diverse perspectives. Most previous literature uses performance indicators that quantify the quality of scientific work produced by a research team, such as the paper's scientific impact, novelty (Funk & Owen-Smith, 2012; Liu et al., 2020; Uzzi, Mukherjee, Stringer, & Jones, 2013), "hit paper" status (Schilling & Green, 2011), and the impact factor of journals in which the paper is published (Garfield, 1999). Scientific impact of publications has traditionally been expressed by citation counts, and can be measured over both the short term and the long term (Bollen, Van de Sompel, Hagberg, & Chute, 2009; Wang, Song, & Barabási, 2013). Short-term and long-term citations are often operationalized as the cumulative number of citations a paper obtained within two or three years (Chai & Menon, 2019; COUNCIL, 2012; Wang et al., 2013), or ten years after its publication (Glänzel & Moed, 2002; Liu, Dehmamy, Chown, Giles, & Wang, 2021; Liu et al., 2018; Zhou & Leydesdorff, 2006), respectively. The journal impact factor is also a popular measure that is used as a proxy for the typical quality of a team's output (Guimera et al., 2005) due to its simple and intuitive definition. Despite some shortcomings and undesirable properties (Bordons, Fernández, & Gómez, 2002; De Bellis, 2009; Seglen, 1997), citations and journal impact factors have been widely used for measuring scientific quality of papers and thus assessing team performance. In this study, we use the cumulative citation count in two years (C2) and ten years (C10) to measure papers' short-term and long-term citations. When analyzing the relationship between freshness and C2, only papers published in and before 2016 are included as C2 is not complete for papers published after that year in the MAG dataset. Similarly, when C10 is analyzed, papers published after 2008 are removed. As the long-term citation indicator is a more powerful predictor of papers' total citations and for detecting highly cited papers (Wang et al., 2013; Wang, 2013), our analyses and discussion will place more focus on C10. To ensure that the drawbacks of citations that are used to measure papers' research quality and team performance will not bias the results, we use the impact factors of journals where the paper is published to measure team performance for a robustness check (see Supplementary Note 1 and Table S12).

**Multivariable regression approaches**

To unveil the evolution of team freshness (RQ1), we calculate the average team freshness by year to make comparisons. Then, we use multivariable regression approaches to address RQ2. To control for various confounding factors, this paper incorporated controls that might be related to citations to a regression model, such as team size (Lee et al., 2015), whether or not the paper is internationally collaborative (Wagner, Whetsell, & Mukherjee, 2019), and the number of unique countries the authors are from (Barjak & Robinson, 2008; Hsiehchen, Espinoza, & Hsieh, 2015). We define *is_international*, as a binary variable which shows if the paper has authors from more than two countries, and the number of countries (*n-country*), which indicates the number of unique countries participating in the collaboration. We use natural logs of C2, C10, and control variables in addition to *is_international* for regression analyses. Ordinary least squares are used to estimate the model:

$$C2_i/C10_i = \alpha + \beta_1(Freshness_i) + \beta_2(Freshness_i \times Freshness_i) + \beta_3(Controls) + C_i + Y_t + \epsilon_{it} \quad (2)$$

In which $C2_i/C10_i$ measures the natural logarithm transformed citation counts (i.e., ln(C2+1) and ln(C10+1)) of paper *i* in a time window of two or ten years; $Freshness_i$ quantifies team freshness of paper *i*; *Controls* represents a set of control variables that are related to citations, i.e., the natural log of the number of authors in a paper (*teamsize*), whether or not paper *i* is internationally collaborative (*is-international*), and the natural log of the number of countries authors are from (*n_country*). To control the time-invariant and discipline-invariant factors, fixed effects regarding publication year ($Y_t$) and those regarding disciplines ($C_i$) are included in Equation 2. The relationship between freshness and citations is not necessarily linear because high levels of freshness might hamper team performance. Therefore, we also consider a possible non-linear association between freshness and citations and incorporate the square term of freshness (Freshness × Freshness) in Equation 2.

To solve RQ3, we divide all publications into separate groups based on team size and perform the above regression analyses again to show the association between team freshness and papers' citations across different levels of team size. The summary statistics and correlation matrix of variables are presented in Table S2 and S3, respectively.

# 4 Results

## 4.1 RQ1: The temporal evolution of team freshness in the past half-century

Generally, team freshness increased over time during the study period. Team freshness has grown over time from 1970 to 2000, remained stable in the 2000s, and increased again after 2010 (see Figure 2(a)). The average team freshness reached 0.443 in 2018, which means that nearly 44% of author pairs in a team have no prior collaboration, on average. To investigate the extreme cases of team freshness, we estimate the proportion of papers with freshness of value 0 and that of value 1. Team freshness of value 0 means that all author pairs in a team have prior collaboration, while team freshness of value 1 indicates that authors in a team have not collaborated with each other before. We find that the proportion of papers with freshness of value 1 has decreased significantly over time, which suggests that teams in which all team members are new to each other have become increasingly rare in modern science. The proportion of papers with freshness of value 0 has remained stable at a low level, less than 20%. On the one hand, teams tend to involve more fresh members, as evidenced by the growing average team freshness by years. On the other hand, the extreme case of freshness, i.e., teams where all members are fresh to each other, is becoming rare. This evidence indicates that scientific teams are inclined to keep a balance between team freshness and repeat collaboration. The increasing team size over time may affect the trend of team freshness. To confirm the upward trend of team freshness, we plot the average team freshness within a given level of team size. The results indicate that small (team size: 2-4), medium (team size: 5-8), and large teams (team size >8) all witnessed a continuous growth of team freshness over time (see Figure S2).

We observe prominent disciplinary differences in the evolution of team freshness. Generally, team freshness in humanities and social sciences is smaller than that in other categories (see Figure 2 (b)). Notably, team freshness in medicine and biology is far larger than that in other categories, reaching around 0.5. This means that, in scientific teams in these two disciplines, nearly 50.1% of author pairs have no prior collaboration. Almost all the disciplines have team freshness lower than 50% (see Figure S3), which means that repeat collaboration still dominates. There is no significant difference in mean growth rates of team freshness in each category, ranging between 17.8% and 17.6%. Almost all the disciplines increased in team freshness since 1970, though with different levels of growth rates (see Figure S4). High average annual growth rates of freshness were found in material science, geology, biology, geography, and chemistry, evidenced by the comparatively steeper slopes of the mean team freshness. In contrast, team freshness in medicine, physics, and computer science did not grow as sharply. It should be noted that medicine has had high team freshness since the 1970s, and remained steady during the study period. This might suggest that scientific teams might deliberately adjust team freshness to keep a balance between freshness and repeat collaboration.

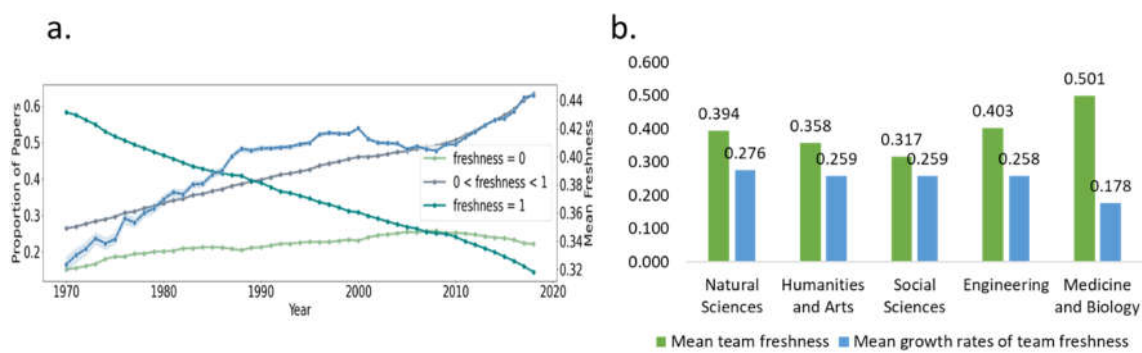

**Figure 2. Average freshness by years and by categories.** In subfigure a, the left Y-axis indicates the proportion of papers in three categories with different levels of freshness, and the right Y-axis refers to average annual team freshness. The blue line indicates average freshness over the years. Subfigure b indicates mean team freshness and mean growth rates of team freshness in five categories of disciplines.

## 4.2 RQ2: An inverted-U-shaped relationship between team freshness and team performance

Because of the coexistence of promoting and detrimental effects of freshness as discussed earlier, we hypothesize that there should be a balance between team freshness and repeat collaboration that brings optimal team performance. Controlling for various confounding factors and introducing the square term of freshness, the multivariable regression models show an inverted-U-shaped association between freshness and papers' C10. Freshness (coefficient: 0.122; p-value<0.001) is significantly positively related to C10 and its square term (coefficient: -0.082; p-value<0.001) is significantly negatively associated with C10 (see Column 1 of Table 1 and Figure 5(a)), indicating a clear inverted-U-shaped relationship between freshness and papers' citations in a ten-year time window. This indicates that incorporating more team freshness brings higher C10 until team freshness reaches a turning point of 29.72%. The inverted-U shape also holds for C2 (see column 2 of Table 1 and Figure 5(b)).

We further observe that, in general, the shape regarding the association between freshness and C2/C10 holds for different time periods and disciplines. The inverse U-shaped curve is found for C2 (see Figure S5 and Table S4) and C10 (see Figure 6 and Table S5) in all the time periods. Additionally, in most disciplines, the inverted-U shape holds for the association between team freshness and papers' C10 (see Figure 7, and tables S8 and S9) and C2 (see Figure S6, and tables S6 and S7). The optimal team freshness for the highest C10 ranges from 8.83% (i.e., mathematics) to 45.91% (i.e., history).

**Table 1. The estimated relationship between freshness and papers' C10 and C2**

|  | (1) | (2) |
|---|---|---|
| Variables | C10 | C2 |
| *Freshness* | 0.212*** | 0.137*** |
|  | (0.004) | (0.002) |
| *Freshness × Freshness* | -0.358*** | -0.327*** |
|  | (0.004) | (0.002) |
| *Team size* | 0.243*** | 0.244*** |
|  | (0.001) | (0.001) |
| *Is_international* | -0.066*** | -0.072*** |
|  | (0.003) | (0.001) |
| *N_country* | 0.584*** | 0.565*** |
|  | (0.006) | (0.003) |
| Field fixed effect | Included | Included |
| Year fixed effect | Included | Included |
| Constant | 1.601*** | 0.573*** |
|  | (0.004) | (0.002) |
| Observations | 12,802,634 | 22,079,726 |
| R-squared | 0.128 | 0.140 |

Notes: Fixed effects regarding year and discipline are incorporated; we used natural logs of the values of C2, C10, and control variables in addition to *is_international* for regression analyses; standard errors in parentheses; *** $p<0.01$, ** $p<0.05$, * $p<0.1$.

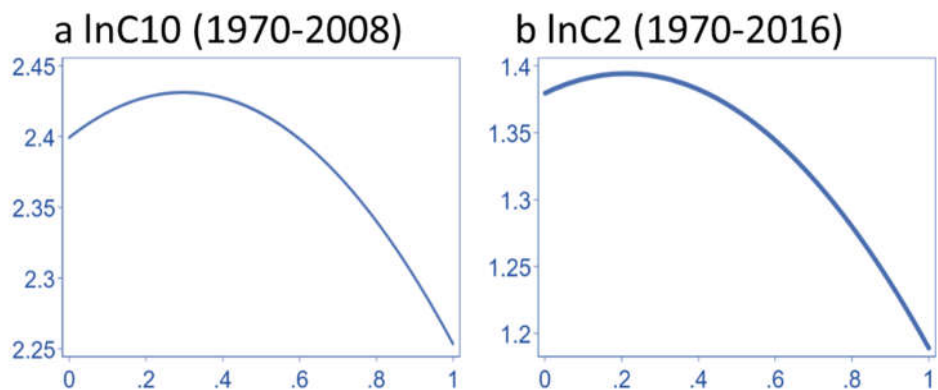

**Figure 3. The estimated association between freshness and papers' C10 and C2.** X-axis indicates team freshness and Y-axis refers to papers' C10 or C2. The shaded area represents ± 1.96 * std. error of each point estimate.

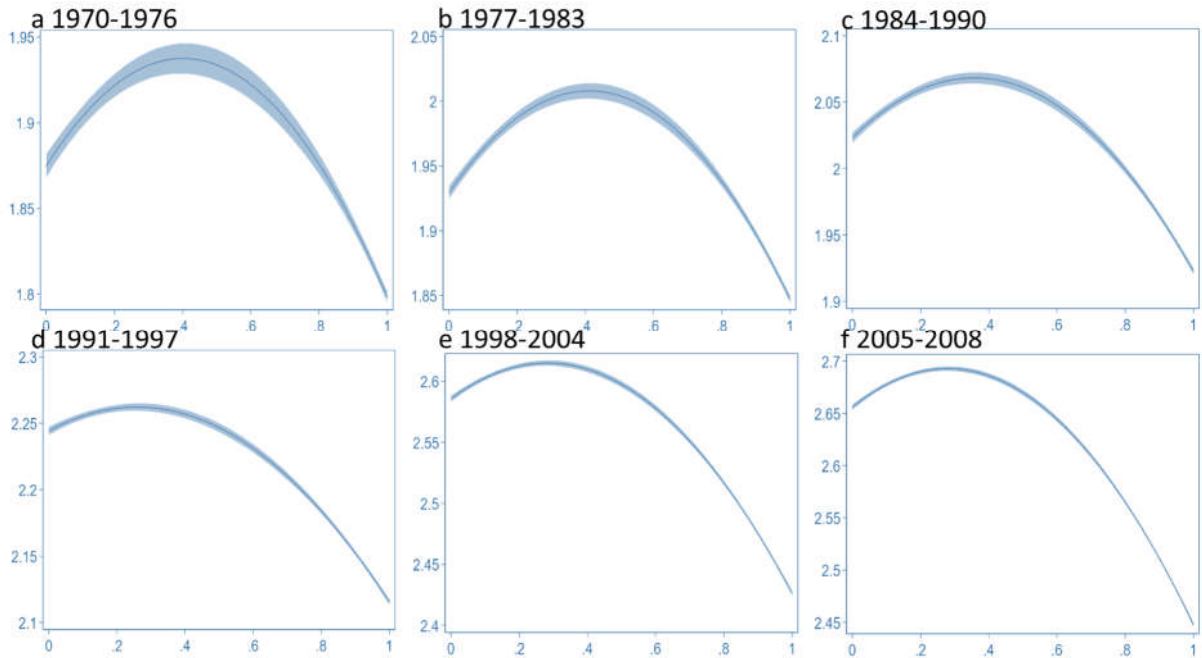

**Figure 4. The estimated association between freshness and papers' C10 by periods from 1970 to 2008.** X-axis indicates team freshness and Y-axis refers to papers' C10. The shaded area represents ± 1.96 * std. error of each point estimate.

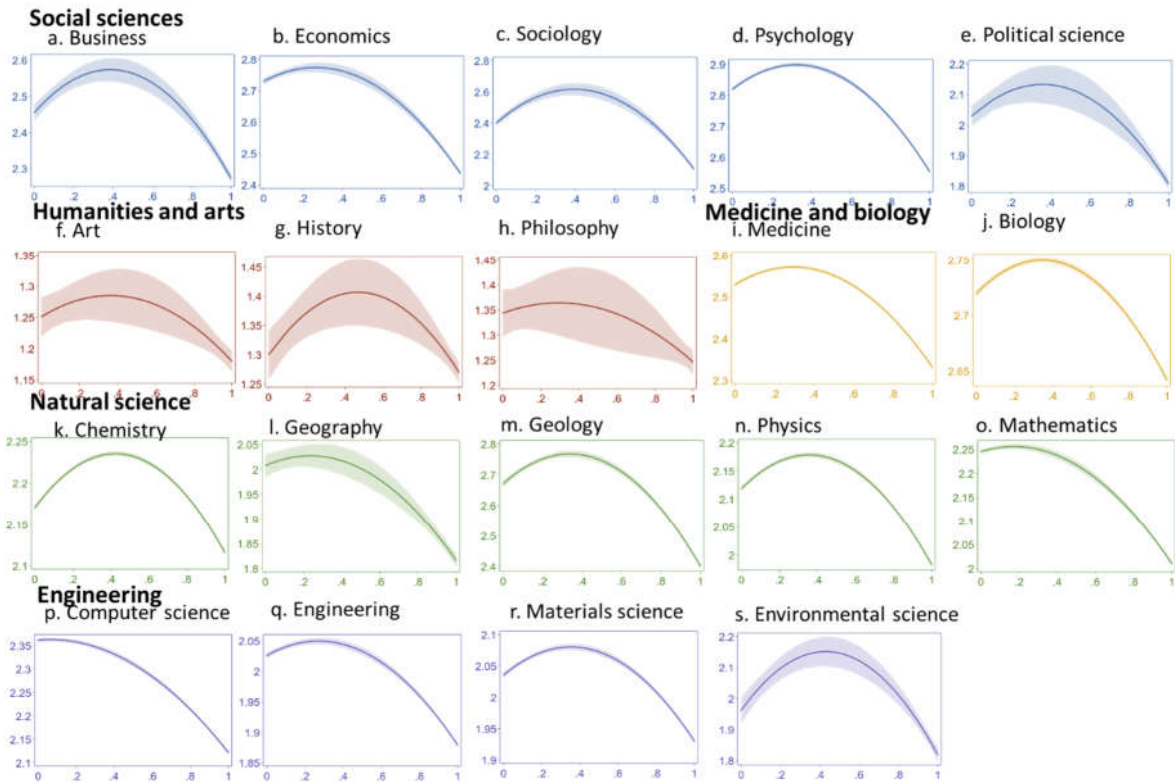

**Figure 5. The estimated association between freshness and papers' C10 by disciplines from 1970 to 2008.** Subfigures (a) to (e) refer to disciplines in social sciences; subfigures (f) to (h) indicate disciplines in humanities and arts; subfigures (i) and (j) indicate the disciplines of medicine and biology, respectively; subfigures (k) to (o) indicate disciplines in natural science; subfigures (p) to (s) indicate

disciplines in engineering. X-axis indicates team freshness and Y-axis refers to papers' C10. The shaded area represents ± 1.96 * std. error of each point estimate.

## 4.3 RQ3: The association between freshness and team performance depends on team size

The relationship between team freshness and team performance varies with changes in team size. The results of multivariable regression models show that controlling for a variety of influential factors, in teams with fewer than ten authors, an inverse U-shaped curve exists for the relationship between team freshness and C10 (see Figure 8 and Table S11), while this shape turns into a positive linear association for teams that include more than nine authors. This evidence suggests a transformation about the shape concerning how team freshness is related to team performance across team size. This finding is consistent for C2 (see Figure S7 and Table S10).

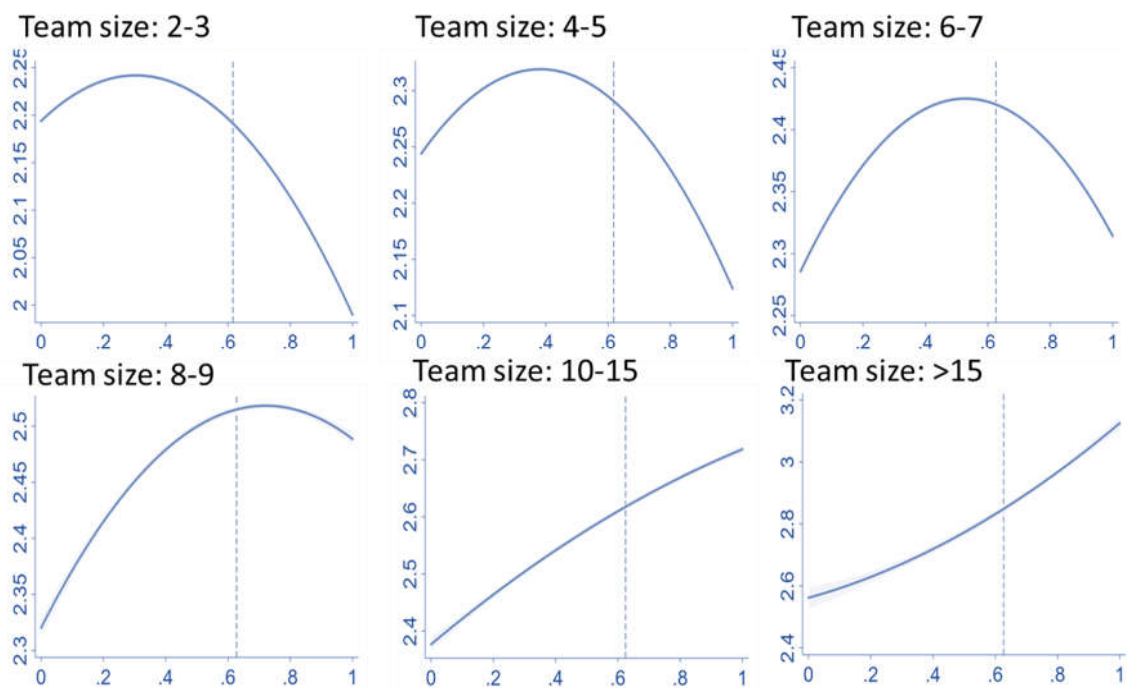

**Figure 6. The estimated association between freshness and papers' C10 by team size from 1970 to 2008.** The dashed line indicates the average team freshness in the sampled papers within a certain level of team size. X-axis indicates team freshness and Y-axis refers to papers' C10. The shaded area represents ± 1.96 * std. error of each point estimate.

# 5 Discussions and Conclusion

Newcomers and old members bring freshness and experience to teams, respectively. Based on more than 43 million research articles published between 1951 and 2018 from the MAG dataset, we extend a measurement that quantifies freshness in teams, investigate the temporal evolution of freshness, and explore its correlation to papers' citations and how this association changes with the variations in team size.

On average, we observe that team freshness is growing over time,[2] while repeat collaboration is still dominant. Involving new members who have no prior collaboration with other members in teams entails significant costs of communication and coordination. In the early years, the construction of a new collaboration relationship was not easy because of the high costs of transportation and communication without advanced information technology. With the decreasing costs of long-distance travel and the development of information and communication technology (Catalini, Fons-Rosen, & Gaulé, 2020), it is easier for scientists to build up new collaboration with others. Therefore, increasing freshness is observed in scientific teams over time. In addition, the historical changes in science provide a foundation for the growing freshness in scientific collaboration, such as the increasing complexity of scientific problems (Katz & Martin, 1997), and the need to combine diverse knowledge and skills to address research problems (Jones, 2009). These changes might make scientists seek new collaborators outside their existing collaboration networks to acquire more diverse and new knowledge that is beyond their existing collaboration networks, and thus increase freshness in teams.

Repeat collaboration still dominates in almost all the disciplines as, apart from medicine, fresh members are the minority in teams. Due to the coexisting advantages and disadvantages of freshness and repeat collaboration, freshness and repeat collaboration should be capped with a certain range to maintain a balance. This might account for the trend we observe that papers with a freshness of 1 are decreasing constantly over time. Furthermore, it could explain why the growth rate of team freshness in medicine is very low, as team freshness has been at a very high level since the 1970s and it cannot grow infinitely. Prominent disciplinary differences are further found in the mean team freshness and the growth of team freshness in the study period. Team freshness in social sciences, and humanities and arts is lower than that in other categories. Team freshness in medicine and biology is the highest among the disciplines, reaching around 50%. This could be explained by the labor division in these two disciplines, where the skills of students or post-doctoral fellows are quite replaceable and thus the entry of fresh members can hardly worsen teams' performance.

We further find an inverted-U-shaped association between freshness and papers' citations across all the periods and disciplines. At the initial stage, more team freshness leads to higher citations. However, after team freshness reaches 29.72%, a higher level of team freshness is associated with a decrease in papers' long-term citations. In other words, when the fraction of new collaboration in a team account for 30% and accordingly repeat collaboration account for 70%, the paper produced by the team will reach its best long-term citations. This result is consistent with the claim in prior studies that the combination of team freshness and repeat collaboration brings the best team performance (Guimera et al., 2005; Perretti & Negro, 2007).

Subsequently, we observe that, with the increase in team size, the inverted-U-shaped relationship between team freshness and citations turns positive. The links between authors are simpler in small teams in which the negative impacts caused by a high level of team freshness will probably hamper team performance. Therefore, there is an optimal level of team freshness. In contrast, in large teams, links generated by newcomers are limited and the disadvantages caused by team freshness can barely influence the whole team. Team members can benefit more from an increasing level of team freshness. Accordingly, we find a positive association between team freshness and papers' citations in large teams. This is consistent with previous claims that a high rate of turnover is important for the survival of large teams, while small teams benefit from the stability of team members (Palla et al., 2007; Whitfield, 2008).

The findings of this study provide implications for the practice of team formation and team management in science. With the increase in the temporal workforce caused by the expansion in the number of awarded PhD degrees and that of academic positions (Milojević, Radicchi, & Walsh, 2018), the incorporation of newcomers is inevitable and thus becomes crucial for team performance. To achieve peak performance, teams should maintain a balance between team freshness and repeat collaboration by containing a majority of incumbents who have existing collaboration with each other and introducing a small proportion of new members. The growing trend of team freshness has been observed in recent decades and the average team freshness, which was around 44% in 2018, has already exceeded the optimal team freshness (i.e., 30%) we find and might even continuously grow in the future. It is more important to retain repeat collaboration rather than introduce newcomers because too much team freshness is harmful to team performance, especially for small teams. Disciplinary differences are prominent concerning the optimal team freshness, which should be considered in the stage of team formation. More importantly, teams should adjust team freshness based on team size because the optimal team freshness for small teams and large teams is substantially distinctive. Additionally, teams, especially small teams, should maintain smooth and efficient communication and coordination between new members and incumbents to alleviate the negative effects caused by a high level of team freshness.

---

[2] The growing trend of team freshness is observed regardless of the value of the parameter we consider for identifying the prior collaboration between authors (see robustness check in Supplementary Note 1).

This study has several limitations. Scientific teamwork is complex and multidimensional, while this study only uses a collection of authors represented on the byline of a publication as a proxy for scientific team. Co-authorship data fails to capture collaboration activities of scientists who are not listed on a research paper. Scientific teamwork that did not lead to publications and scientists who made contributions but were not listed as authors are not considered in this study. In this sense, survey, interview, and experimental data should be included to study scientific teamwork more directly and show how team freshness influences other forms of scientific teamwork. Moreover, team performance could be reflected in a variety of dimensions, such as team efficiency and team novelty. This study only investigates the quality indicators of papers, i.e., scientific impacts of papers produced by teams, and the impact factor of journals where papers are published. More team performance indicators should be considered for a better understanding of how team freshness influences team outcomes. Despite the multiple dimensions of team freshness, such as freshness of team members, we only focus on fresh relations among team members. Team freshness measured by team members' career age or past productivity will be analyzed for future study. In addition, we only include a few confounding factors because other influential factors of team performance are not easily captured and controlled. Therefore, this study reveals a correlation rather than a causality between team freshness and team performance. Whether or not there remains a causal relationship deserves deeper investigation. Heterogeneities concerning culture and policy environments might exist for the relationship between team freshness and team performance, as the formation and management of scientific teams could be largely shaped by these two factors. Whether and the extent to which differences in culture and policy environments reshape the association between team freshness and team performance will be studied subsequently.

# Supplementary Information

Supplementary Information (SI) is available for this paper:

https://www.dropbox.com/s/l83exeetn4ozxnk/supplementary%20information37.docx?dl=0

# Acknowledgements

This work is supported by the Youth Program of National Natural Science Foundation in China (No: 2104054 and 72104007), the Youth Project of Humanities and Social Sciences of the Ministry of Education (MOE) of China (No: 21YJC870001) and the Youth Project of Shanghai Science and Technology Innovation Action Plan Soft Science (21692196000). The authors deeply appreciate the constructive comments from the reviewers.